\documentclass[letterpaper, 10 pt, conference]{ieeeconf}  
\IEEEoverridecommandlockouts                             

\usepackage{amssymb}
\usepackage{makecell}
\usepackage{amsmath}
\usepackage{color,xcolor}
\usepackage{algorithm, algpseudocode}

\usepackage[utf8]{inputenc}
\usepackage{booktabs, caption, makecell}

\usepackage{threeparttable}
\usepackage{bbm}
\usepackage{tikz}
\usetikzlibrary{positioning}
\usepackage{multirow}
\usepackage[figuresright]{rotating}
\usepackage{amsmath}

\usepackage{tablefootnote}

\usepackage{subcaption}

\usepackage{tikz-network}
\usepackage{algorithm, algpseudocode}

\usepackage{multirow}

\title{ \LARGE \bf Approximate Dynamic Optimization via Deep Neural Operators}

\author{Amin Nassaji$^{1}$, Ilias Mitrai$^{2}$, Prodromos Daoutidis$^{1}$
\thanks{This work was supported by the McKetta Department of Chemical Engineering (IM) and NSF-CBET Award No. 2313289 (PD)}
\thanks{$^{1}$Amin Nassaji and Prodromos Daoutidis are with the Department of Chemical Engineering and Materials Science, University of Minnesota, Minneapolis, MN, USA {\tt\small \{nassa033@umn.edu, daout001@umn.edu\}}}
\thanks{$^{2}$Ilias Mitrai is with the McKetta Department of Chemical Engineering, The University of Texas at Austin, 78712, Austin, TX, USA
        {\tt\small imitrai@che.utexas.edu}}}

\begin{document}

\maketitle
\thispagestyle{empty}
\pagestyle{empty}

\begin{abstract}

This paper addresses the solution of nonlinear dynamic optimization problems that compute optimal manipulated input profiles to enforce desired output profiles. Such trajectory optimization problems commonly arise in chemical process applications, for example, batch processes where optimal temperature or feeding profiles (in case of fed-batch processes) are calculated to enforce time-varying product quality profiles, tightly controlling the reaction rate or rate of heat generation. 
We propose deep neural operators that approximate function to function mappings as surrogates for the solution of such dynamic optimization problems.  
We specifically employ deep operator networks (DeepONets) and Fourier-enhanced DeepONets in a batch polymerization reactor case study for which number-average and weight-average molecular weight profiles, together with a final conversion target, are enforced through an optimal temperature program. 
Our results show that the Fourier-enhanced DeepONet architecture performs very well in approximating the solution of the dynamic optimization problem for different instances, achieving a lower prediction error compared to the standard DeepONet architecture and standard feedforward neural networks.

\end{abstract}

\section{Introduction}

Dynamic optimization (DO) refers to the class of optimization problems that involve time-varying decisions and dynamical systems \cite{biegler2010}. It is at the heart of numerous process operation and control problems, for example, in batch processes that are inherently dynamic, or in systems subject to frequent transitions in operating conditions \cite{soroush1999review}.

In DO problems, one typically determines an optimal time-varying trajectory of the manipulated inputs (flow rates, heating/cooling loads, pressure) to optimize a criterion that can be economic (maximizing profit) or performance related (minimizing batch time, maximizing conversion, yield or selectivity, etc.) \cite{betts2010,biegler2010}. A particularly important and challenging class of DO problems involves trajectory optimization. In such problems, the target is a trajectory rather than a single value of the output variable. Examples include batch polymerization reactors that operate at a time-varying temperature to enforce a desired product quality profile linked to the desired properties (molecular weight distribution, conversion, etc.),
fed-batch bioreactors, control problems in batch processes that aim to enforce a precomputed optimal trajectory of manipulated inputs (for example, an optimal feeding or heating/cooling strategy), and process startup or shutdown where the process needs to follow a specific operating path for safety reasons \cite{haugwitz2009start}.
Another major class involves transient operations and load changes, including grade transitions in polymerization and specialty chemicals plants, where constraints must be continuously satisfied as the process moves from one product grade to another (often specified by allowable property ranges) \cite{soroush1999review}, and the goal is to find feasible trajectories that avoid unsafe regions while minimizing transition time and off-spec production \cite{mahadevan2002grade,shi2016gradetransition}.

In all of these applications, the key point is that performance and feasibility depend on the entire time evolution of the system, not just the endpoint. These optimization problems often need to be solved for different target profiles, different initial conditions, and different time horizons, so each variation triggers a new optimization solution which can quickly become a computational bottleneck in online decision-making or large scale design and operation studies \cite{biegler2010,betts2010}. This motivates approaches that reduce the need to solve a full dynamic optimization problem from scratch for each instance and instead reuse the computation or learned structure across instances \cite{karg2020efficient,kumar2021industrial, mitrai2025discovering}. However, many such approaches are developed primarily for finite-dimensional optimization, where an instance is represented by a fixed-length vector and the solution is a finite set of decision variables. In contrast, in trajectory optimization the solution map is an operator $\mathcal{G}$ over functions, i.e., it connects an input profile $p(t)$ to a control trajectory $u(t)$, written as $u(t)=\mathcal{G}(p)(t)$.
This motivates operator learning, where a surrogate model learns the mapping from an input function (e.g., a desired profile or scenario description) to an output function (the corresponding optimal trajectory) \cite{kovachki2024handbook}.

\begin{figure*}
    \centering
    \includegraphics[scale=1.05]{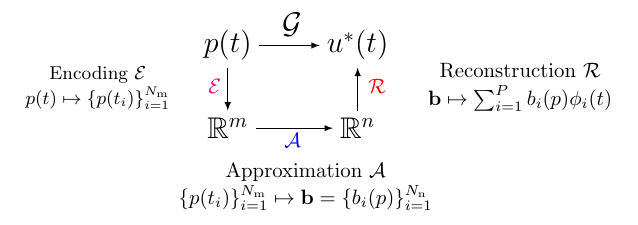}
    \caption{Illustration of the three steps of operator learning frameworks.}
    \label{fig: deeponet concept}
\end{figure*}

Data-driven operator learning has received significant attention in the literature recently \cite{kovachki2024handbook, wang2021pideeponet,hwang2022aaai,wang2021selfsup,krstic2024neuralop,wang2025backstepping,lamarque2025adaptive, seidman2022nomad}. In general, operator learning frameworks have three steps: encoding, approximation, and reconstruction (see Figure~\ref{fig: deeponet concept}). The encoding step creates an embedding of the input function into a finite-dimensional space. This is usually achieved by discretizing the input function at specific time points. This step accounts for the fact that, even though the input is a function, in practice it is sampled at discrete time points. The approximation step captures the effect of the operator in the discrete embedded space. The reconstruction maps the discrete space prediction to the continuous space, enabling the prediction of the operator's output at any time point.

Although some theoretical aspects of operator learning have long been studied \cite{chen1995universal}, recent advances have focused on new deep learning architectures with guaranteed approximation properties. Specifically, deep operator networks (DeepONets) \cite{lu2021deeponet} have been used to approximate (or learn) the solution of partial differential equations for the simulation and control of PDE-constrained systems and in backstepping methods \cite{hwang2022aaai,wang2021selfsup,krstic2024neuralop,wang2025backstepping,lamarque2025adaptive}. DeepONets use feedforward neural networks for the approximation and reconstruction steps. These networks are known as branch and trunk networks.

This prior body of work motivates a key distinction in how we propose to use deep neural operators in this paper. In existing applications, the operator approximates the system simulator (or an inverse map) and is then embedded inside an optimizer or estimator \cite{wang2021pideeponet,wang2021selfsup,hwang2022aaai}. In contrast, our goal is to learn the optimization solution operator itself: a direct map from a problem specification (targets, initial conditions, constraints, and horizon information) to near-optimal control trajectories consistent with the constrained optimizer’s solution.

In this paper, we focus on the application of DeepONets to approximate the solution of trajectory optimization problems, especially those arising in the operation of batch process systems. To improve the accuracy of standard DeepONets, we extend the input to the trunk network with Fourier features, which allow the architecture to learn profiles over long time horizons \cite{zhu2023fourierdeeponet}. We apply these DeepONet architectures to a case study in the operation of a batch polymerization reactor for which typical product quality objectives such as conversion, number-average molecular weight, weight-average molecular weight, and polydispersity can be achieved through an optimal temperature program. For such systems, optimal temperature programs often feature a mid-batch valley and late ramp as the process moves from kinetic to diffusion-limited regimes; this makes the dynamics stiff and the optimal profile computation challenging \cite{ahn1998}. Our results show that the Fourier-enhanced architecture achieves a lower prediction error over a wide range of test scenarios compared to the standard DeepONet architecture and standard feedforward neural networks.

The remainder of the paper is organized as follows. Section II presents the application of DeepONets for approximating the solution of trajectory optimization problems. Section III describes the batch polymerization case study. Section IV describes the data generation and training procedure. In Section V, we discuss the results and analyze the prediction accuracy of the different models. Finally, Section VI provides conclusions of this work.

\section{Operator learning for Trajectory optimization}
\subsection{Trajectory optimization problems}
We consider a dynamical system with states $x(t)\in\mathbb{R}^n$ and manipulated variables $u(t)\in\mathbb{R}^m$. The dynamic behavior of the system is described by an ordinary differential equation
\begin{equation}
\dot{x}(t)=f(x(t),u(t)) \label{eq:dyn}
\end{equation}
Given a desired trajectory profile $p(t)$ (e.g., a setpoint or specification over time), we seek the optimal input trajectory $u^*(p)(t)$ by solving the trajectory optimization problem

\begin{equation}\label{eq:ocp}
\begin{aligned}
u^{*}(p)(\cdot) \in \arg \min_{u(\cdot)} \ \ &
\Phi\!\big(x(T_f),T_f,p\big) \\
&\quad + \int_{0}^{T_f} \ell\!\big(x(t),u(t),p(t),t\big)\,dt \\
\text{s.t.} \ \ &
\dot{x}(t) = f\!\big(x(t),u(t)\big) \\
& u_{\min} \leq u(t) \leq u_{\max}
\end{aligned}
\end{equation}
\noindent
where $\ell(\cdot)$ denotes the running (trajectory) cost, and $\Phi(\cdot)$ denotes a terminal cost (e.g., terminal deviation penalties and/or a penalty on $T_f$). $u_{\min},u_{\max}\in\mathbb{R}^{n_u}$ are lower/upper limits for manipulated variables.
We follow a direct solution approach to solve the optimal control problem. We convert the continuous-time problem into a finite-dimensional NLP by representing $u(t)$ with a finite-dimensional smooth parameterization. The manipulated input is specified at the $N$ time points on $[0,T_f]$, and smooth interpolation between these points yields a continuous control $u(t;\theta)$, where $\theta\in\mathbb{R}^N$ stores the values at these time points. With this parameterization, simple lower/upper input limits become simple lower/upper bounds on the components of $\theta$. The rate limit is modeled by two linear inequalities for each adjacent pair of discretization points,
\begin{equation}
-\bar r\,\Delta t_k \;\le\; u_{k+1}-u_k \;\le\; \bar r\,\Delta t_k,
\qquad k=1,\ldots,N-1
\label{eq:rate}
\end{equation}

Given $\theta$, we propagate the state by integrating \eqref{eq:dyn} from $t=0$ to $t=T_f$ under $u(t;\theta)$. If numerical integration fails or produces non-finite states, we return a large objective value for that $\theta$ so that the optimizer steers away from that region. The optimizer first estimates gradients, then selects a direction \(d^{(k)}\) that will decrease the objective while reducing constraint violations. It then performs a line search to choose a step size \(\alpha^{(k)}\) and updates the parameters as $\theta^{(k+1)}=\theta^{(k)}+\alpha^{(k)} d^{(k)}$. This parameterization of the input together with the constraints yields a smooth constrained NLP in \(\theta\). The optimizer then searches over \(\theta\) to solve this constrained NLP. Constrained gradient-based sequential quadratic programming methods are well suited for smooth constrained NLPs with simple upper and lower limits on the decision variables, linear constraints that bound changes between adjacent discretization points, and smooth nonlinear terminal constraints \cite{nocedalwright2006}. Because the system dynamics are nonlinear, the resulting NLP is generally nonconvex. In nonconvex problems, different initial guesses can lead the solver to different locally optimal solutions. A practical strategy is multi-start initialization: create several feasible control profiles that already satisfy bounds and rate limits and run the NLP from each one. Because feasibility is enforced at the time points, each starting point integrates stably while exploring a different part of the search space. We keep the best feasible result across runs based on the objective, which improves robustness without increasing the number of decision variables \cite{nocedalwright2006}.

After finding an optimal control \(u^{*}(t)\) for one specification (set of input parameters), we re-optimize the system and record the time series for the time \(t\), the optimal input \(u^{*}(t)\), and the outputs \(p(t)\) for different specifications. This produces a data set that maps each specification to its optimal input trajectory \(u^{*}(t)\) together with the associated output trajectories \(p(t)\).

\subsection{Approximate Trajectory optimization via DeepONets}

The solution of the dynamic optimization problem presented above is a mapping between two functions, $p(t)$ and $u^{*}(t)$. Thus, we are interested in learning the map $\mathcal{G}$
\begin{equation}
\mathcal{G}:\; P \to U \qquad u^{*}(t) = \mathcal{G}(p)(t)
\label{eq:operator}
\end{equation}
where $U$ and $P$ are the spaces of the input and output functions. 

\begin{figure}
    \centering
    \includegraphics[scale=0.75]{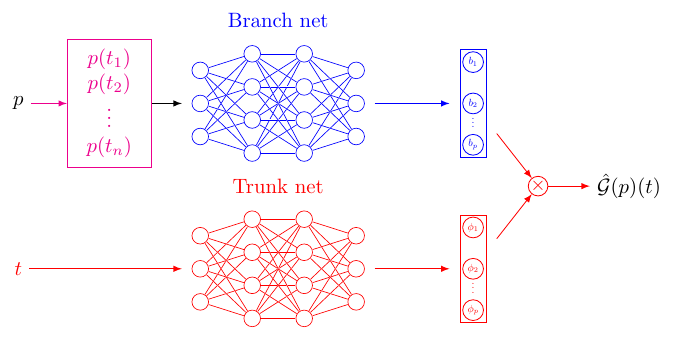}
    \caption{The DeepONet architecture. The magenta panel lists sampled inputs \(p(t_i)\). Blue nodes form the branch net that encodes the sampled inputs into a coefficient vector \(b\); red nodes form the trunk net that encodes the query \(t\) into features \(\phi(t)\). The crossed circle denotes the output, computed as the dot product \(b(p)^\top \phi(t)=\hat{\mathcal{G}}(p)(t)\).}
    \label{fig:deeponet fig}
\end{figure}

We approximate the operator $\mathcal{G}$ with a DeepONet, denoted $\hat{\mathcal{G}}(p)(t)$, which has two components (see Fig.~\ref{fig:deeponet fig}). The branch network $b(p;w_{p})\in\mathbb{R}^p$ encodes the target trajectory $p$ from its values in a fixed set of sampling points, and the trunk network $\phi({t};w_{\phi})\in\mathbb{R}^p$ encodes the query coordinate $t$. $w_{p}$ and $w_{\phi}$ are the weights of the branch and trunk networks, respectively. The predicted value at $t$ is then given by the dot product 
\begin{equation}
\hat{u}(t) = \hat{\mathcal{G}}(p)(t) = b(p)^\top \phi({t}) \label{eq:deeponet}
\end{equation}

The branch and trunk weights ($w_{b}, w_{\phi}$) are learned by minimizing the per-point prediction error over the training dataset set consisting of samples of input signals $p(t)$, the associated $u^{*}$ and a set of query time points. Overall, the training data is $\mathcal{D} = \{p_{i},\;\{(t_{ij},\,u(t_{ij})\}_{j=1}^{N_{J}}\}_{i=1}^{N_{I}}$, where index $i$ represent the trajectory and index $j$ the sampling time in trajectory $i$. $N_{I}$ and $N_{J}$ are the number of trajectories and sample points within each trajectory, respectively. 

In our model, the trunk network takes a query time $t$ and a set of sine / cosine time features. These Fourier features help the model capture gradual changes and oscillations. We form the time feature vector from normalized time $\tilde{t}=t/T_f$ as
\begin{equation}
\phi(\tilde{t})
=
\big[
\,\tilde{t},\ 
\{\sin(2\pi k\,\tilde{t})\}_{k=1}^{K},\ 
\{\cos(2\pi k\,\tilde{t})\}_{k=1}^{K}
\big]^{\!\top}
\end{equation}

\section{Case Study: Batch Polymerization}
We consider batch polymerization of methyl methacrylate (MMA) in solution (see \cite{ahn1998} for details on the process and modeling equations). The manipulated input is the reactor temperature $T(t)$, limited by bounds and a rate limit for safe and practical operation. The desired output variables are conversion $X$, number-average molecular weight $M_n$, weight-average molecular weight $M_w$, and polydispersity $PD$. These can be calculated using standard moment relations:
\begin{align}
X(t) &= 1 - \frac{M(t)}{M(0)} \label{eq:X}\\[6pt]
M_n(t) &= \frac{W_m\,[\,G_1(t)+F_1(t)\,]}{G_0(t)+F_0(t)} \label{eq:Mn}\\[6pt]
M_w(t) &= \frac{W_m\,[\,G_2(t)+F_2(t)\,]}{G_1(t)+F_1(t)} \label{eq:Mw}\\[6pt]
PD(t) &= \frac{M_w(t)}{M_n(t)} \label{eq:PD}
\end{align}
where $W_m$ is the monomer molecular weight and $M(t)$ denotes the moles of monomer remaining in the reactor at time $t$ (with $M(0)$ the initial charged moles). The symbols $I$ and $S$ denote the initiator and solvent, respectively. The quantities $G_0,G_1,G_2$ are the zeroth, first, and second moments of the live-radical chain-length distribution, and $F_0,F_1,F_2$ are the corresponding moments of the dead-polymer chain-length distribution. The initial conditions are the charged amounts $I(0),M(0),S(0)$; polymer moments start at zero unless the seed is present. The total volume $V(t)$ changes with the composition and temperature through density correlations, so even in a pure batch the effective concentrations change as the polymer forms.

The dynamic model couples material balances and moment balances with temperature-dependent kinetics (initiator decomposition, propagation/termination, and chain transfer) and composition/temperature effects through free-volume based rate modifiers; we refer to \cite{ahn1998,brandrup1999} for the full governing equations, parameter definitions, and density correlations. These relations capture the gel/glass regime, where rising viscosity limits radical mobility and lowers termination rates, often producing a mid-batch valley in optimal temperature profiles as the system moves from kinetic to diffusion-limited behavior \cite{ahn1998}. Additional assumptions include a well-mixed batch with no spatial gradients, no feed addition (pure batch operation), and a temperature trajectory $T(t)$ that an external heater can enforce. We omit the energy balance, considering the temperature as a directly controlled manipulated input, negating the need to model the heat transfer equipment \cite{ahn1998,brandrup1999}.

\subsection{Implementation: Models, Data, and Training}

\subsubsection{Data generation and representation}
Each training case corresponds to one solved instance of the trajectory optimization problem in Section~II-A. The specification includes two desired product-quality trajectories, $M_n(t)$ and $M_w(t)$, together with scalar final conversion $X(T_f)$.
For each case, we solve the dynamic optimization problem to obtain the optimal temperature trajectory $T^{*}(t)$ using an SQP-based constrained optimizer (SLSQP) \cite{kraft1988slsqp}.

We work on normalized time $\tau=t/T_f\in[0,1]$ and represent all trajectories on a uniform grid of $N_t$ points in $\tau$. The DeepONet branch input is formed by combining the sampled trajectories and scalars,
\[
p_i \;=\;\big[\,M_{n,i}(\tau_1{:}\tau_{N_t}),\;M_{w,i}(\tau_1{:}\tau_{N_t}),\;X_i(T_f)\big],
\]
and the supervised output is the optimal temperature $T_i^{*}(\tau)$ evaluated at query points $\tau\in[0,1]$. 

\subsubsection{Train/test splits and experimental settings} 
We generate a dataset of $N$ solved optimal control instances and reserve a fixed $20\%$ of cases as an unseen test set. This same test set is reused across all subsequent experiments so that any change in performance is due to the experimental change (e.g., less training data or horizon truncation), not because different cases were tested.

To evaluate data efficiency, we train each model on different fractions of the available training cases. To evaluate generalization across time horizons, each case $i$ is assigned a horizon fraction $r_i\sim \mathrm{Uniform ([R_{\min},R_{\max}])}$, and training uses only samples satisfying $\tau\le r_i$. For evaluation, we compute error metrics only on the same truncated interval $\tau\le r_i$ for each test case. This setting reflects variable horizon trajectory requirements where we may only need the portion of the trajectory up to the current horizon, and it highlights the advantage of coordinate-based operator models when horizons vary.

\subsubsection{Neural model classes and architectures}
We train four networks: two Plain DeepONets (a smaller one and a wider one), one Fourier DeepONet, and a fully connected feedforward neural network (FNN) baseline. The point of having two Plain DeepONets is to separate the effect of increased parameter count from improved time features. Plain DeepONet~B tells us what happens if we simply make the network bigger, while Fourier DeepONet tells us what happens if we keep the parameter count comparable but give the trunk a richer (Fourier) representation of time. Finally, the FNN baseline provides a non-operator machine-learning (ML) reference, so we can see what we gain by using the DeepONet function to function structure instead of a conventional vector to vector predictor.

For the DeepONet models, the branch network encodes the discretized specification vector $p_i$ into a coefficient vector $b(p_i)\in\mathbb{R}^{q}$, while the trunk network encodes the query coordinate $\tau$ into $\phi(\tau)\in\mathbb{R}^{q}$, and the prediction is
\[
\widehat{T}(\tau)=b(p_i)^{\top}\phi(\tau)
\]

In the Fourier DeepONet, the trunk input is augmented with sine/cosine features of normalized time \cite{zhu2023fourierdeeponet,tancik2020fourierfeatures}. The FNN baseline maps the same discretized input vector $p_i$ directly to the discretized temperature vector $T^{*}_i(\tau_1{:}\tau_{N_t})$ on the fixed grid.

\subsubsection{Normalization, loss functions, and training}
All input channels are standardized using statistics computed on the training set only. Trajectory inputs are normalized per channel across the training cases, and scalar inputs are normalized accordingly. The output temperature is centered using the training-set mean.

Models are trained with the Adam optimizer (learning rate $10^{-3}$) for a fixed number of iterations \cite{kingma2015adam}. The neural-operator models minimize a pointwise regression loss over supervised case time pairs, while the FNN baseline minimizes an error over the full output vector on the fixed grid. We choose FNN baseline sizes so that the total parameter counts of Plain DeepONet, Fourier DeepONet, and the FNN baseline are of comparable magnitude to ensure a fair comparison. All models were implemented in DeepXDE \cite{lu2021deepxde}.

\section{Results}

In all experiments, we compare each model’s predicted temperature trajectory
$\widehat{T}(t)$ against the optimizer solution $T^{*}(t)$ on the same time grid, and we report trajectory-level errors aggregated over time points and test cases. We emphasize three settings: (i) full-horizon prediction on representative test cases, (ii) performance as training data are reduced (data efficiency), and (iii) robustness when predicting trajectories over variable time horizons.

\subsection{Full-horizon prediction on unseen test cases (fixed test set)}
We first train all four models using a fixed split with 80\% of cases for training and a fixed 20\% unseen test set for evaluation. This same test set is reused across all subsequent experiments so that any change in performance is due to the experimental change (e.g., less training data or horizon truncation), not because different cases were tested.

\begin{figure}[!t]
  \centering
  \begin{subfigure}{\linewidth}
    \centering
    \includegraphics[width=\linewidth]{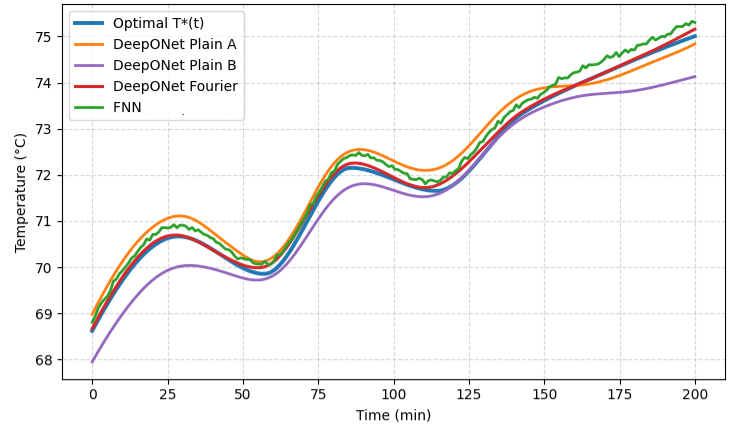}
  \end{subfigure}\hfill
  \begin{subfigure}{\linewidth}
    \centering
    \includegraphics[width=\linewidth]{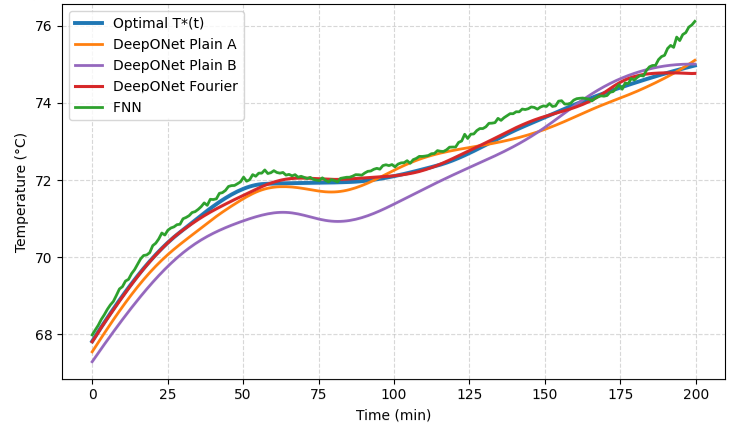}
  \end{subfigure}
  \begin{subfigure}{\linewidth}
    \centering
    \includegraphics[width=\linewidth]{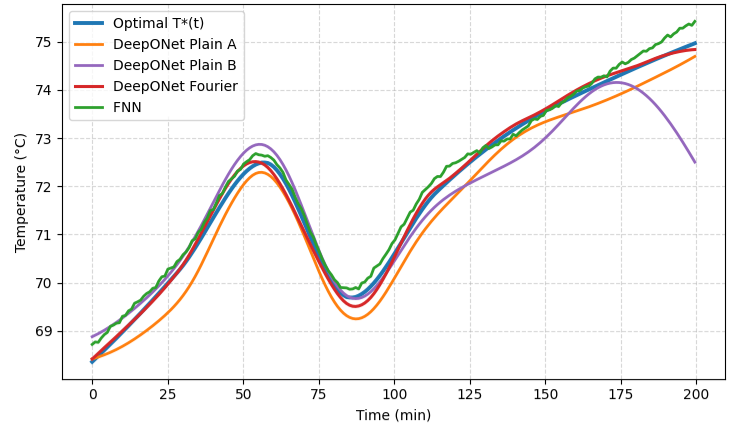}
  \end{subfigure}
  \caption{Three test examples, comparing the optimizer solution $T^{*}(t)$ with predictions from Plain DeepONet~A, Plain DeepONet~B (wider), Fourier DeepONet, and the FNN baseline.}
  \label{fig:case1_examples_full}
\end{figure}

Figure~\ref{fig:case1_examples_full} considers three representative unseen test examples, comparing the optimizer solution $T^{*}(t)$ against the predictions from Plain DeepONet~A, Plain DeepONet~B (wider), Fourier DeepONet, and the FNN baseline.
In these examples, the Fourier DeepONet tracks the optimizer temperature trajectory closely over the full horizon, reproducing the main shape features most notably the mid-batch valley with consistent timing and slope. Simply increasing the width of the plain DeepONet (Plain DeepONet~B) does not reliably improve the prediction, even though it has more parameters, and the smaller plain model (Plain DeepONet~A) tends to be the most sensitive to shape changes. The Fourier DeepONet also outperforms the FNN baseline in these examples, highlighting the benefit of DeepONet’s function-to-function formulation for trajectory prediction.

\subsection{Data efficiency: performance vs.\ training fraction}
To see how performance changes when we have fewer labeled training cases, we retrain the same four model classes using smaller training sets while keeping the same fixed 20\% test set. Specifically, we retrain the models using reduced training fractions of 60\% and 40\% of the total dataset (with all training cases always drawn from the non-test pool). Table~\ref{tab:dataeff_case1} reports mean absolute error (MAE) and mean squared error (MSE), averaged over all test cases.

Reducing training data hurts all models, but not equally. The Fourier DeepONet performs better in every training fraction, and its error increases more slowly as we reduce the amount of training data. The FNN baseline is reasonably close when training data are plentiful (80\%), but  deteriorates noticeably at 60\% and 40\%, indicating weaker data efficiency. The two plain DeepONets degrade the most at 40\%, and the wider plain model (Plain~B) does not show a consistent advantage over the smaller plain model (Plain~A). In other words, simply increasing the size of the plain model does not consistently resolve its errors; enriching the time input with Fourier features improves accuracy more consistently without a large increase in parameters.

\begin{table}[t]
\centering
\caption{Full-horizon test-set performance under reduced training fractions with a fixed 20\% test set.
Errors are averaged over all test cases.}
\label{tab:dataeff_case1}
\begin{tabular}{llcc}
\toprule
\textbf{Setting} & \textbf{Model} & \textbf{MAE} & \textbf{MSE} \\
\midrule
\multirow{4}{*}{Train 80\% / Test 20\%}
& Plain DeepONet~A & 0.1501 & 0.1458 \\
& Plain DeepONet~B (wider) & 0.1564 & 0.1196 \\
& \textbf{Fourier DeepONet} & \textbf{0.0812} & \textbf{0.0543} \\
& FNN baseline & 0.1019 & 0.0655 \\
\midrule
\multirow{4}{*}{Train 60\% / Test 20\%}
& Plain DeepONet~A & 0.1556 & 0.1448 \\
& Plain DeepONet~B (wider) & 0.1928 & 0.1738 \\
& \textbf{Fourier DeepONet} & \textbf{0.0756} & \textbf{0.0477} \\
& FNN baseline & 0.1425 & 0.1029 \\
\midrule
\multirow{4}{*}{Train 40\% / Test 20\%}
& Plain DeepONet~A & 0.2909 & 0.2749 \\
& Plain DeepONet~B (wider) & 0.2740 & 0.3189 \\
& \textbf{Fourier DeepONet} & \textbf{0.1430} & \textbf{0.1027} \\
& FNN baseline & 0.1809 & 0.1384 \\
\bottomrule
\end{tabular}
\end{table}

\subsection{Moving-horizon generalization: random-horizon truncation}
One of the main advantages of coordinate-based operator models is that they can be evaluated over different time horizons without changing the model architecture. To test this directly, we train and evaluate each model under random-horizon truncation. This setting mirrors moving-horizon implementation, where we may only need the portion of the trajectory up to the current horizon. Table~\ref{tab:randh_case1} reports moving-horizon (random-horizon truncation) performance in terms of MAE and MSE. The Fourier DeepONet remains the most accurate model under random horizons, while the FNN baseline deteriorates substantially, providing a complementary evaluation in a variable-horizon setting.

\begin{table}[t]
\centering
\caption{Random-horizon test-set performance when each case is truncated to a random horizon
$r_i \in [0.5,1.0]$.}
\label{tab:randh_case1}
\begin{tabular}{lcc}
\toprule
\textbf{Model} & \textbf{MAE} & \textbf{MSE} \\
\midrule
Plain DeepONet~A & 0.1548 & 0.1527 \\
Plain DeepONet~B (wider) & 0.1717 & 0.1315 \\
\textbf{Fourier DeepONet} & \textbf{0.0808} & \textbf{0.0590} \\
FNN baseline & 0.2336 & 0.2545 \\
\bottomrule
\end{tabular}
\end{table}

\section{Conclusions}

This paper proposed and evaluated the use of deep neural operators to approximate the solution of dynamic optimization (DO) problems, specifically for trajectory optimization. In such problems, the solution map is an operator that takes an input specification function (desired product quality profiles and scalar case parameters) and returns an optimal manipulated-input trajectory. We learn this function to function map directly using DeepONet architectures.

We considered a batch polymerization reactor as a case study, where a temperature program is used to enforce trajectories of conversion and molecular-weight targets. The system is highly nonlinear and the optimizer solution often exhibits a mid-batch valley and a late ramp as the process moves from kinetic to diffusion-limited regimes. We trained four surrogate models, Plain DeepONet~A (smaller), Plain DeepONet~B (wider), a Fourier-enhanced DeepONet, and a fully connected feedforward neural network (FNN) baseline and evaluated them on a fixed unseen test set. The results show that enriching the trunk time coordinate with Fourier features improves full-horizon trajectory matching and remains the most accurate and data-efficient model as training data are reduced. Under random-horizon truncation (i.e., evaluating each case only up to a randomly selected horizon) the Fourier DeepONet retains strong performance, highlighting a practical advantage of coordinate-based operator models for variable-horizon prediction.

Beyond open-loop use, the same surrogate can be deployed in closed-loop form by repeatedly updating the problem specification online and re-evaluating the operator to produce an updated control segment.


\addtolength{\textheight}{-12cm}   

\vspace{10pt}
\addcontentsline{toc}{section}{References}
\bibliographystyle{ieeetr}
\bibliography{refs}


\end{document}